\documentclass[]{aa} 

\usepackage{dirtytalk}
\usepackage{natbib}
\usepackage{graphicx}
\usepackage{epsfig}
\usepackage{txfonts}
\usepackage{hyperref}
\usepackage{url}
\usepackage{xcolor}
\hypersetup{
colorlinks,
linkcolor={red!80!black},
citecolor={blue!80!black},
urlcolor={blue!80!black}
}

\def\cm3{cm$^{-3}$}
\def\kms{km~s$^{-1}$}
\def\msunyr{M$_{\odot}$\,yr$^{-1}$}

\def\msun{M$_{\odot}$}

\def\two{\ts {\,\sc ii}}
\def\three{\ts {\,\sc iii}}

\def\beq{\begin{equation}}
\def\eeq{\end{equation}}

\def\lesssim{\mathrel{\hbox{\rlap{\hbox{\lower4pt\hbox{$\sim$}}}\hbox{$<$}}}}
\def\gtrsim{\mathrel{\hbox{\rlap{\hbox{\lower4pt\hbox{$\sim$}}}\hbox{$>$}}}}

\def\two{{\,\sc ii}}
\def\three{{\,\sc iii}}

\def\ip{\rho}

\def\v1d{{\tt V1D}}

\def\cmfgen{{\tt CMFGEN}}
\def\longpol{{\tt LONG\_POL}}
\def\heracles{{\tt HERACLES}}

\def\aj{AJ}

\def\apj{ApJ}
\def\apjs{ApJS}
\def\apjl{ApJL}
\def\aap{A\&A}
\def\araa{ARA\&A}

\def\mnras{MNRAS}

\linenumbers

\begin{document}

   \title{Spectropolarimetric modeling of interacting Type II supernovae.
   Application to early-time observations of SN\,1998S.}

   \titlerunning{Modeling of the SN\,1998S polarization.}

\author{
   Luc Dessart\inst{\ref{inst1}}
  \and
   Douglas C. Leonard\inst{\ref{inst2}}
  \and
  Sergiy S. Vasylyev\inst{\ref{inst3},\ref{inst4}}
  \and
   D. John Hillier\inst{\ref{inst5}}
}

\institute{
Institut d'Astrophysique de Paris, CNRS-Sorbonne Universit\'e, 98 bis boulevard Arago, F-75014 Paris, France\label{inst1}
  \and
    Department of Astronomy, San Diego State University, San Diego, CA 92182-1221, USA\label{inst2}
      \and
  Department of Astronomy, University of California, Berkeley, CA 94720-3411, USA\label{inst3}
  	\and
	Steven Nelson Graduate Fellow in Astronomy\label{inst4}
	  \and
    Department of Physics and Astronomy \& Pittsburgh Particle Physics, Astrophysics, and Cosmology Center (PITT PACC),  University of Pittsburgh, 3941 O'Hara Street, Pittsburgh, PA 15260, USA\label{inst5}
}
   \date{}

\abstract{
  High-cadence surveys of the sky are revealing that a large fraction of red-supergiant (RSG) stars, which are progenitors of Type II-Plateau (II-P) supernovae (SNe), explode within circumstellar material (CSM). Such SNe II-P/CSM exhibit considerable diversity, with interaction signatures lasting from hours to days, potentially merging with the Type IIn subclass for which longer-duration interaction typically occurs. To tackle this growing sample of transients and to understand the pre-SN mass loss histories of RSGs, we train on the highest quality, spectropolarimetric observations of a young Type IIn SN taken to date:  Those of SN 1998S at $\sim$\,5\,d after explosion. We design an approach based on a combination of radiation hydrodynamics with \heracles\ and polarized radiative transfer with \cmfgen\ and \longpol. The adopted asymmetries are based on a latitudinal, depth- and time-independent, scaling of the density of 1D models of SNe II-P/CSM (e.g., model r1w6b with a ``wind'' mass-loss rate of 0.01\,\msunyr\ used for SN\,2023ixf). For a pole-to-equator density ratio of five, we find that the polarization  reaches, and then remains for days, at a maximum value of 1.0, 1.4, and 1.8\,\%  as the CSM extent is changed from 6, to 8 and 10\,$\times$\,10$^{14}$\,cm. The polarization is independent of wavelength away from funnel-shaped depolarizations within emission lines. Our models implicate a significant depolarization at line cores, which we use to constrain the interstellar polarization of SN\,1998S. Our 2D, prolate ejecta models with moderate asymmetry match well the spectropolarimetric observations of SN\,1998S at 5\,d, supporting a polarization level of about $\sim$\,2\,\%. This study provides a framework for interpreting future spectropolarimetric observations of SNe II-P/CSM and SNe IIn and fostering a better understanding of the origin of their pre-SN mass loss.
}


\maketitle


\section{Introduction}
\label{sect_intro}

 There is much interest today about the origin of pre-explosion mass loss and subsequent circumstellar material (CSM) that surrounds the massive star progenitors that lead to Type II supernovae (SNe). Some rare and extraordinary super-luminous SNe II exhibit signatures of ejecta interaction with CSM for weeks or months (e.g., SN\,2010jl; \citealt{zhang_10jl_12}; \citealt{fransson_10jl}; \citealt{D15_2n}), suggesting extreme CSM properties of several \msun\ that extend out to many 10$^{15}$\,cm, and earning these events a classification as Type IIn SN \citep{niemela_etal_85,schlegel_iin_90}. In contrast standard-luminosity Type II SNe, with more typical Plateau or fast-declining light curves, exhibit signatures of interaction for only a few days after the emergence of the shock at the progenitor surface before showing the more standard spectral properties of noninteracting Type II SNe with Doppler-broadened lines \citep{bruch_csm_21,bruch_csm_22,wynn_pap2_24} -- we may refer to this second class of events as SNe II-P/CSM. From this growing sample of events emerges a finer diversity, with objects like SN\,2013fs exhibiting interaction, IIn-like signatures for about a day \citep{yaron_13fs_17}, and others such as SN\,1998S for one or two weeks \citep[hereafter L00]{leonard_98S_00}.

There are likely multiple origins for this diversity of CSM and SN properties, which may involve wave excitation from the stellar core \citep{quataert_shiode_12,fuller_rsg_17}, nuclear flashes \citep{WH15}, surface pulsations \citep{yoon_cantiello_rsg_14}, or binary effects \citep{wu_fuller_22}. Red-supergiant (RSG) star atmospheres may also be more massive and extended than typically envisioned, serving as a wasteland for the numerous instabilities taking place in the stellar envelope or at its surface and acting over the entire RSG-phase duration (see, e.g., \citealt{d17_13fs}; \citealt{soker_csm_21}; \citealt{fuller_tsuna_rsg_24}). Some of these phenomena may occur in unison.

Characterizing this massive-star mass loss is a prerequisite for understanding its nature, both in terms of CSM density and extent. New insights about the dynamical properties of the CSM are starting to be revealed with high-cadence high-resolution observations at the earliest post-breakout times \citep{shivvers_98S_15,smith_23ixf_23,pessi_24ggi_24}. Constraining the geometry of the CSM may also provide clues on the mechanism at the origin of the CSM. For unresolved sources, polarization is a powerful means to find evidence for asymmetry  and constrain its nature \citep{wang_wheeler_rev_08}.

Spectropolarimetric observations of interacting SNe have been secured for only a few objects such as SN\,2010jl \citep{patat_10jl_11} or SN\,2009ip \citep{mauerhan_pol_09ip_14}. In Type II SNe with early-time signatures of interaction present for only a few days, the spectropolarimetric observations must be conducted at the earliest times after shock breakout when a large part of the CSM remains unshocked. This has been achieved recently for SN\,2023ixf \citep{vasylyev_23ixf_23,singh_23ixf_pol_24}, but undoubtedly the best quality, early-time, spectropolarimetric observations of a SN with signatures of interaction are those obtained for SN\,1998S by L00 at an estimated post-explosion time of 5\,d. Being of relatively high spectral resolution (i.e., 6\,\AA), these data reveal a polarization that varies significantly from continuum to emission line regions, as well as across line profiles between the broad wings and the narrow cores. While much uncertainty and debate surround the correction for interstellar polarization (ISP; see. e.g., L00 or \citealt{wang_pol_01}),  SN\,1998S clearly shows percent level intrinsic polarization that indicates significant departures from spherical symmetry.

Converting this intrinsic polarization into a specific CSM asymmetry is difficult. The presence of electron-scattering wings on all emission lines indicates that the continuum and lines form under optically-thick conditions \citep{chugai_98S_01,d09_94w} -- estimates requiring optically-thin conditions (see., e.g., \citealt{Brown_McLean_77}) cannot be used. Interpretations connecting an overall polarization level to a specific degree of asphericity have been formulated \citep{hoeflich_87A_91} but applied to noninteracting SNe like SN\,1987A where the ejecta structures are vastly different from those present in interacting SNe (see also \citealt{DH11_pol}). In such phenomena, modeling of both the radiation-hydrodynamics and the radiative transfer are necessary, as done in a first attempt by \citet{D15_2n} and applied to the spectropolarimetric observations of the Type IIn SN\,2010jl \citep{patat_10jl_11}.

In this letter, we extend our previous work on the polarization modeling of noninteracting Type II SNe during the photospheric \citep{dessart_12aw_21} and nebular phases \citep{dessart_pol_blob_21,leonard_hv_pol_21} by considering the cases of interacting Type II SNe. The essence of the approach (our ``ansatz"), which is simple, is to enforce homologous expansion throughout the interaction region in order to use the 2D polarized radiative-transfer code \longpol\ \citep{hillier_94,hillier_96, DH11_pol,dessart_12aw_21} in which homologous expansion is currently assumed. As we show in the next section, our ansatz is well motivated at early times, but could also be used at all times if only the continuum polarization is sought.  In the next section, we present our modeling procedure, followed by our results in Section~\ref{sect_res} and a comparison of our models to the spectropolarimetric observations of SN\,1998S in Section~\ref{sect_obs}. While the presentation is kept concise and focused in the main text, we present the code \longpol\ in Appendix~\ref{sect_nomenclature}. All simulations in this work will be uploaded on zenodo.\footnote{\url{https://zenodo.org/communities/snrt}}


\begin{figure}
\centering
\includegraphics[width=\hsize]{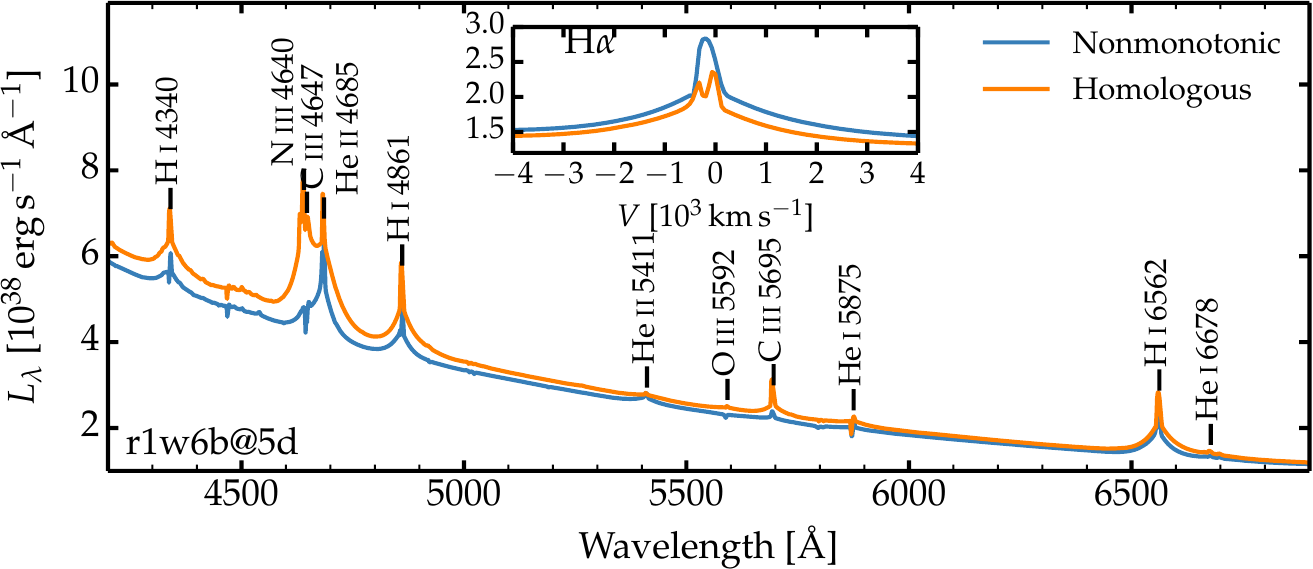}
\caption{Comparison of the optical luminosity of model r1w6b at 5\,d after explosion and computed either with the nonmonotonic solver or assuming homologous expansion and using the blanketed mode in \cmfgen. [See discussion in Section~\ref{sect_setup}.]}
\label{fig_spec_comp}
\end{figure}

\section{Numerical procedure}
\label{sect_setup}

 In this work, we use offshoots of the model r1w6, which itself was one instance in the grid of interacting, SNe II-P/CSM models from \citet{d17_13fs}. It corresponds to a solar-metallicity progenitor star of 15\,\msun\ on the zero-age main sequence, whose explosion following core collapse is designed to produce an ejecta with a kinetic energy of $1.2\,\times\,10^{51}$\,erg. This explosion occurs in a CSM corresponding to a wind with a mass loss rate of 0.01\,\msunyr, a terminal velocity $V_\infty$ of 50\,\kms\ and a velocity profile versus radius $R$ given by $V(R)=V_\infty (1 - R_\star/R)^\beta$ with $\beta=$\,2 ($R_\star$ is the progenitor surface radius). The dense part of the CSM extends to $R_{\rm CSM}$, beyond which the mass-loss rate drops smoothly and within a few 10$^{14}$\,cm to 10$^{-6}$\,\msunyr. This configuration is modeled in 1D with the radiation-hydrodynamics code \heracles\ \citep{gonzalez_heracles_07} as discussed in \citet{d17_13fs}. For a selection of epochs, the \heracles\ calculations are post-processed with the 1D radiative transfer code \cmfgen\ \citep{HD12,D15_2n} in order to generate for each model the variation of the electron density, opacity, and emissivities versus radius. During the \cmfgen\ calculation, the gas temperature is kept fixed and equal to the value from the \heracles\ snapshot.

 In our modeling of the polarization, the asymmetry of the interaction is considered only at this last stage,  by introducing a latitudinal scaling of the densities, opacities and emissivities, which are then used by the 2D, polarized radiative transfer code \longpol\ \citep{hillier_94,hillier_96,DH11_pol,dessart_12aw_21}. This approach is not fully-consistent hydrodynamically but it is flexible and allows for explorations of different asymmetries. In practice, we use a simple latitudinal scaling of the density that goes as $X(\mu)=a(1+A\mu^2)$, where $\mu=\cos\theta$, $\theta$ is the polar angle, $A$ takes values typically of a few, and $a$ is chosen to preserve the density at a given depth. We solve the 2D polarized radiative transfer for such 2D ejecta from 3800 to 9500\,\AA. In the 2D ejecta, the electron density is scaled by $X$, whereas the opacities and the emissivities are scaled by $X^2$ (for full details of the method and an application to the modeling of spectropolarimetric observations of Type II-P SN\,2012aw, see \citealt{dessart_12aw_21}; see also Appendix~\ref{sect_nomenclature}).

In \citet{d17_13fs}, all radiative-transfer calculations were done with the 1D steady-state, nonmonotonic solver in \cmfgen\ in order to capture the complex interaction structure and model the spectral evolution in detail. Unfortunately, \longpol\ does not currently handle nonmonotonic flows -- homologous expansion is required. To circumvent this current limitation of \longpol, we modify the velocity from the \heracles\ snapshot that is read in by \cmfgen, and enforce homologous expansion by setting $V=R/t$ and $t=R_{\rm phot} / V_{\rm phot}$. We then run \cmfgen\ with a different solver, in steady-state and 1D, but in blanketed mode \citep{hm98}. A similar blanketed mode in \cmfgen\ was also used by \citet{shivvers_98S_15} in their modeling of high-resolution spectra of SN\,1998S, although without any coupling to radiation hydrodynamics.

Figure~\ref{fig_spec_comp} compares the emergent optical spectra for model r1w6b, for which $R_{\rm CSM}$ is $8 \times 10^{14}$\,cm, at 5\,d using either the original nonmonotonic velocity (and nonmonotonic solver) or assuming homologous expansion (and using the blanketed mode in \cmfgen; see also Appendix~\ref{appendix_cmfgen}). The difference in the optical luminosity is small and the morphology of emission lines is largely preserved -- this confirms that the main broadening mechanism is electron scattering. However, the blanketed mode resolves a discrepancy in the strength of the blend of N\three\ and C\three\ multiplets around 4640\,\AA, which the nonmonotonic solver systematically underestimates relative to observations (see, e.g., \citealt{jacobson_galan_23ixf_23}). These changes in line strength arise in part from the greater UV luminosity obtained with the blanketed mode. Other optical emission lines are of similar strength in both \cmfgen\ calculations and in good agreement with observations of SN\,1998S (see Section~\ref{sect_obs}) or SN\,2023ixf \citep{jacobson_galan_23ixf_23}. This close correspondance between the two \cmfgen\ calculations suggests that our ansatz is acceptable.

In this work, we focus mostly on model r1w6b but also consider 2D ejecta built from variants in which  $R_{\rm CSM}$ is 6 and  10\,$\times$\,10$^{14}$\,cm (models r1w6[a,c]). Assuming homologous expansion, we performed \cmfgen\ calculations for the r1w6[a,b,c] models at times between 1.67--2.5\,d up to 5-10\,d after explosion. These epochs straddle the phase during which the photosphere is located in the unshocked CSM and eventually into the cold-dense shell (for a general overview of this evolution, see \citealt{dessart_review_24}). In all cases presented here, the 2D ejecta built from these interaction models are prolate and have a moderate pole-to-equator density ratio of five. Furthermore, the variation with latitude as $\mu^2$ is progressive and thus excludes jets and disks.

\begin{figure}
\centering
\includegraphics[width=\hsize]{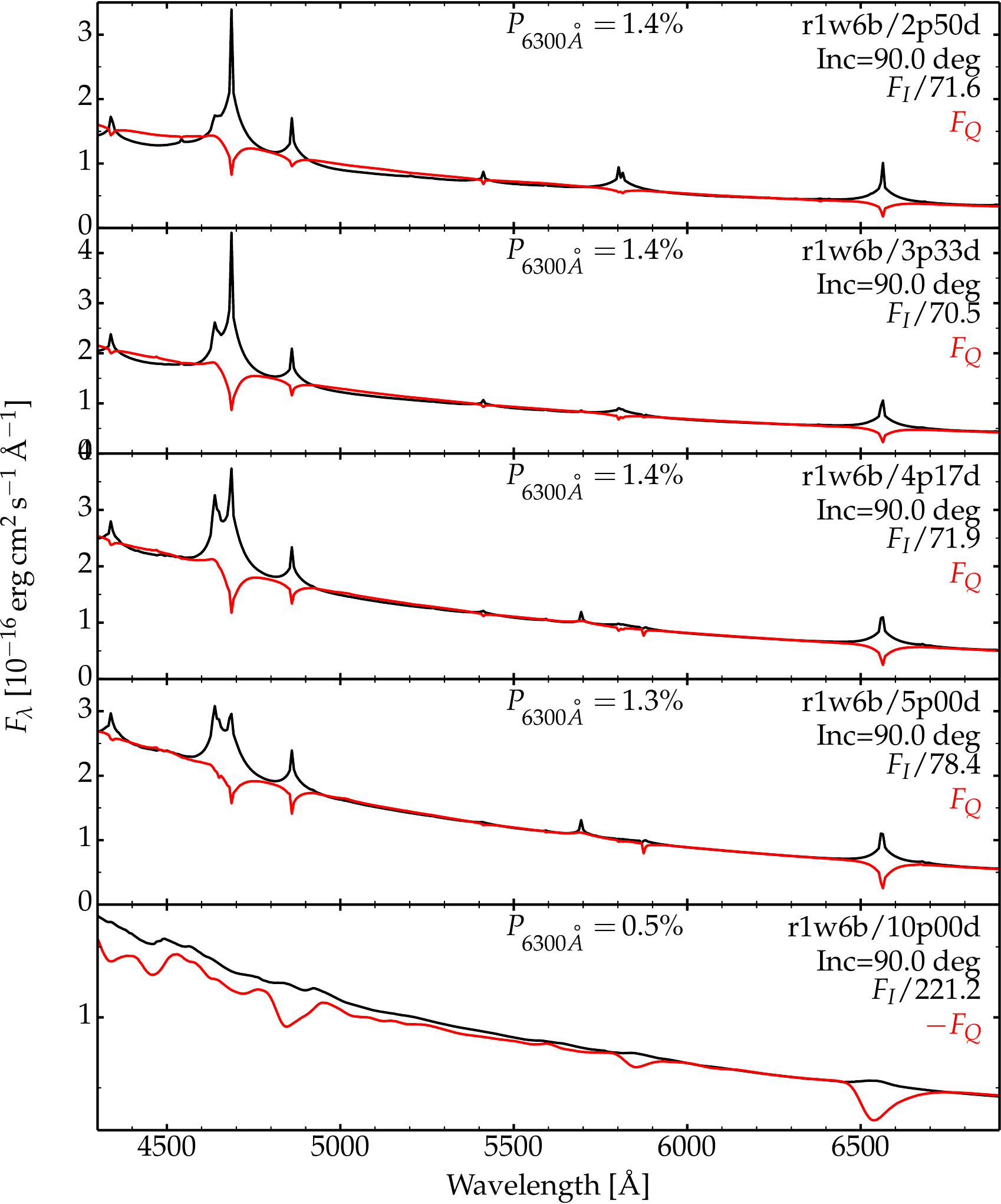}
\caption{Evolution of the total flux $F_I$ and the polarized flux $F_Q$ computed with \longpol\ assuming a 2D, prolate ejecta based on model r1w6b ($\rho_{\rm pole}/\rho_{\rm eq}=$\,5). Epochs cover from 2.5 to 10.0\,d after explosion. Both fluxes are scaled to the inferred distance of SN\,1998S, with an additional scaling for the total flux (see label) to match the magnitude of the polarized flux at 6300\,\AA\ ($F_Q$ flips sign at the last epoch). All fluxes have been rebinned at 6\,\AA. [See discussion in Section~\ref{sect_res}.]}
\label{fig_r1w6b_evol}
\end{figure}

\section{Polarization modeling results}
\label{sect_res}

Figure~\ref{fig_r1w6b_evol} shows the results from the 2D polarized radiative-transfer code \longpol\ for a 2D, prolate ($\rho_{\rm pole}/\rho_{\rm eq}=$\,5) ejecta based on model r1w6b, for a 90-deg inclination, and from 2.5 to 10.0\,d after explosion. The total flux $F_I$ shows emission lines with narrow cores and broad, symmetric, electron-scattering broadened wings for the first four epochs whereas at the last epoch at 10.0\,d, the total flux exhibits weak, blueshifted, Doppler broadened lines that are starting to show blueshifted, P-Cygni absorptions. With the normalization of the total and polarized fluxes (i.e., total flux multiplied by the percent polarization, here denoted as $F_Q$) at 6300\,\AA, one sees that the slope of each is essentially identical. In other words, the overall continuum polarization is constant with wavelength. Across most lines, the polarized flux decreases, with a maximum reduction at the line cores; this indicates substantial depolarization of both line and continuum photons at these wavelengths.

The lower polarization across lines arises in part from the fact that all lines form exterior to the continuum-formation region and thus at lower electron-scattering optical depth (see second panel from top in Fig.~\ref{fig_init}). The linear polarization arising from scattering with free electrons is therefore greater for continuum than for line photons. In the narrow line cores, we see photons that have undergone no frequency redistribution in wavelength (or in velocity space) as they travelled outwards from their original point of emission (many of these ``core'' photons are emitted beyond the photosphere, at low electron-scattering optical depth, and thus could not exhibit much polarization). These line core photons thus experienced little or no scattering with free electrons and are thus unpolarized. Furthermore, because the optical depth in lines such as H$\alpha$ is huge, the (polarized) continuum photons overlapping with the line cores suffer from absorption, leading to a narrow polarization dip in emission line cores.

 The funnel-shaped profile observed in polarized flux across lines indicates that the associated photons are depolarized (i.e., the polarized flux is below the level obtained by interpolating $F_Q$ from the adjacent continuum regions). This funnel shape arises because continuum photons originally emitted at a $\lambda_{\rm init}$ not too far from a line's rest wavelength $\lambda_{\rm c}$ will be absorbed by that line if they ever come within a few Doppler widths (say 10\,\kms) of $\lambda_{\rm c}$. This eventuality may occur as photons random-walk and scatter with free electrons in the CSM. The probability for this is greater the smaller is $|\lambda_{\rm init}$-$\lambda_{\rm c}|$. Line cores act as a sink for neighboring (in $\lambda$-space), scattered continuum photons.

Some subtleties are also visible. For example, the C\three/N\three\ blend exhibits a weaker depolarization. These lines form deeper than H$\alpha$ and thus closer to the region of continuum formation (see also Fig.~\ref{fig_init}, and L00), but also through different processes (i.e., continuum fluorescence and dielectronic recombination). The polarization is found to be positive for the first four epochs, thus aligned with the axis of symmetry. This results from an optical depth effect, suggesting that polarization is controlled primarily by the lower-density, equatorial regions where the bulk of the radiation emerges rather than the higher density regions where scattering is enhanced (see discussion by \citealt{DH11_pol}). The polarization flips in sign at the last epoch (equivalent to a 90-deg rotation of the polarization angle), when the unshocked CSM has been fully swept-up and the entire spectrum forms in the dense shell.

In this 2D, prolate ejecta model with $\rho_{\rm pole}/\rho_{\rm eq}=$\,5, the continuum polarization (which is essentially constant throughout the optical range) has a maximum value of 1.4\,\% until 4.17\,d, dropping to 1.3\,\% at 5.0\,d and 0.5\,\% at 10.0\,d. Hence, the presence of an extended unshocked CSM with a modest asymmetry can produce percent-level polarization at early times (as inferred in SNe\,1998S or 2010jl), without invoking extreme explosion asymmetries (e.g., jets or disks), and thus comparable to peak values obtained at the transition to the nebular phase of non-interacting, Type II SNe  \citep{leonard_iauga_15,nagao_13ej_21}. Modulations of $R_{\rm CSM}$ do not affect the qualitative results obtained for r1w6b but change the values of the maximum polarization and its evolution in time (Appendix~\ref{sect_appendix_figs}). The equivalent 2D prolate ejecta based on model r1w6a (r1w6c) exhibit a maximum polarization of 1.0\,\% (1.8\,\%) and a decline at $\sim$\,5\,d ($\sim$\,10\,d). Consequently, SNe II-P/CSM with longer-lived IIn-like signatures should on average exhibit a greater level of polarization. Other parameters that may impact the polarization behavior are the CSM density or associated wind mass loss.


\begin{figure}
\centering
\includegraphics[width=0.6\hsize]{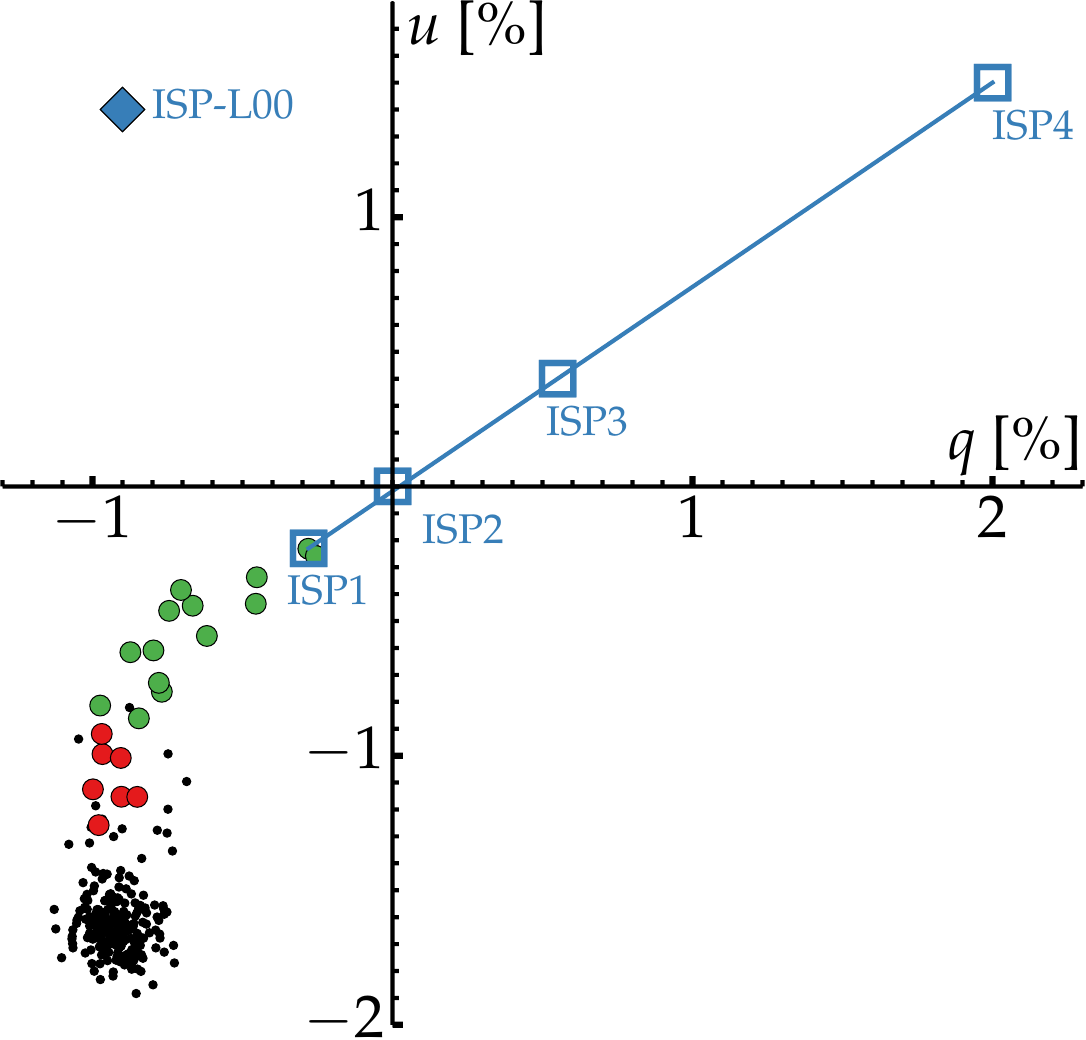}
\caption{Polarization data in the $q-u$ plane for SN\,1998S at $\sim$\,5\,d after explosion, from L00. Photons with wavelengths corresponding to the broad wings and narrow-line core of the H$\alpha$ flux profile are indicated with red and green circles, respectively.  Each point represents a bin 10\,\AA\ wide, except for the green circles, which are binned at 2\,\AA\ for enhanced resolution.  The blue line spans the range of potential ISP choices bounded by the ISP derived from different assumptions about the extent of depolarization in the narrow-line core region of H$\alpha$.  The blue squares indicate four specific choices of ISP discussed in the text, which yield the inferred intrinsic polarizations displayed in the bottom panels of Fig.~\ref{fig_comp_mod_obs}.
[See Section~\ref{sect_obs} for discussion.]}
\label{fig_obs_q_u}
\end{figure}

\begin{figure}
\centering
\includegraphics[width=\hsize]{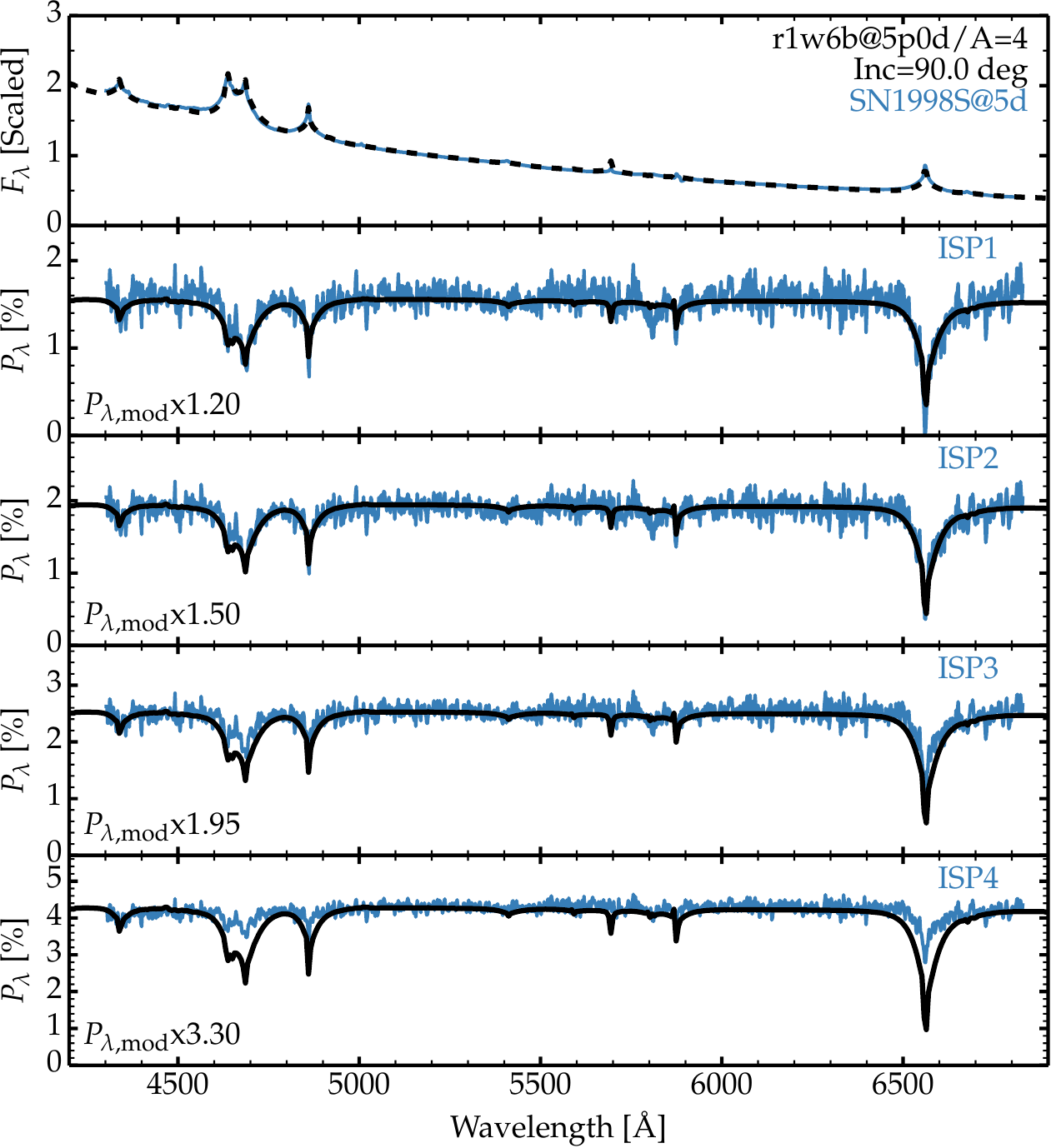}
\caption{Comparison of the spectropolarimetric observations of SN\,1998S at 5\,d after explosion with the counterparts obtained with \longpol\ assuming a 2D, prolate ejecta ($\rho_{\rm pole}/\rho_{\rm eq}=$\,5) based on model r1w6b at 5\,d. The observer's inclination relative to the axis of symmetry is 90\,deg. Four ISP choices are shown, corresponding to ISP[1,2,3,4] indicated in Fig.~\ref{fig_obs_q_u} ; note that ``ISP2'' represents the observed data, with no ISP removed. Observations have been corrected for a recession velocity of 840\,\kms\ and a reddening $E(B-V)$ of 0.11\,mag. Model spectra have been smoothed and rebinned to 6\,\AA\ to match the resolution of the observations. [See Section~\ref{sect_obs} for discussion.]}
\label{fig_comp_mod_obs}
\end{figure}

\section{Comparison to spectropolarimetric observations of SN\,1998S}
\label{sect_obs}

We now compare our modeling results with the single-epoch spectropolarimetry of SN 1998S at $\sim$\,5\,d post-explosion presented and thoroughly discussed by L00. We choose to compare our models with these data since they remain (to our knowledge) the only optical spectropolarimetry obtained both early enough to capture the strong ejecta-CSM interaction and with sufficient resolution ($\sim$\,6\,\AA) to clearly reveal the spectropolarimetric behaviors of both the narrow- and broad-line features in an SN\,IIn.

 The observed data of SN\,1998S from L00 present a high degree ($\sim$\,2\,\%) of wavelength-independent continuum polarization across the observed spectral range (4314--6850\,\AA) and sharp depolarizations across all strong emission lines (see Fig.~\ref{fig_obs_q_u}, as well as the ``ISP2'' panel of Fig.~\ref{fig_comp_mod_obs} here).  To convert these observed data into intrinsic polarization that can be directly compared with our models, we must first determine a plausible value of the ISP.

We present the observed data of SN\,1998S in the Stokes $q-u$ plane in Fig.~\ref{fig_obs_q_u}. In this representation, the black circles clustering around [$q,u$] $\approx$ [-0.9\%,-1.7\%] arise from the continuum regions; the red and green circles correspond specifically to the broad wings and narrow core spectral regions of H$\alpha$.

This presentation enables immediate identification of an interesting fact: The broad wings and narrow core photons ``point'' along different directions in the Stokes $q-u$ plane.  Indeed, L00 used this to derive two different ISP possibilities, one with the assumption of unpolarized photons in the broad wings of strong lines, and one under the assumption of unpolarized photons contributed solely by the narrow emission lines; these choices are indicated by the points identified in Fig.~\ref{fig_obs_q_u} as ``ISP-L00'' and ``ISP4'', respectively.

Here we shall be guided by the modeling results of the previous section, which predict that the spectral region corresponding specifically to the narrow-line core of strong emission features (i.e., H$\alpha$) should consist not only of intrinsically unpolarized narrow-line photons, but also a significantly --- if not completely --- depolarized underlying continuum at these wavelengths.  This presents to us a new range of ISP choices, for which the bounds can be easily established. First, if the narrow-line core of H$\alpha$ is assumed to have zero intrinsic polarization (i.e., unpolarized line photons and completely depolarized underlying continuum), then the ISP must lie simply at the location of the tip of the narrow H$\alpha$ line in the $q-u$ plane; this is indicated by ``ISP1'' in Fig.~\ref{fig_obs_q_u}. At the other extreme lies the result that obtains if we assume only unpolarized narrow-line photons and no depolarization of the underlying continuum photons; this is indicated by ``ISP4'' in Fig.~\ref{fig_obs_q_u}.  The line connecting these two points then spans the range of allowable ISP values under different assumptions about the degree of depolarization of the underlying continuum. ``ISP3'', located at [$q, u$] = [0.55\%, 0.40\%] is one such example, and was chosen specifically to best fit the depolarization at H$\alpha$ found by model r1w6b at 5\,d.  Curiously, the origin is also an allowable ISP choice, and so we also consider that possibility and label that point ``ISP2''.

When the four ISP choices labeled in Fig.~\ref{fig_obs_q_u} are removed from the SN\,1998S data, they yield the intrinsic polarizations shown in the bottom panels of Fig.~\ref{fig_comp_mod_obs}.  While the inferred overall level of continuum polarization changes drastically with ISP choice, it is notable that the basic features of the resulting spectropolarimetry do not: A high level of wavelength-independent continuum polarization with ``funnel-shaped'' depolarizations across the strong line features. The level of depolarization across lines is best matched by choices ISP[1,2,3], which imply an intrinsic continuum polarization in SN\,1998S of $\sim$\,2\,\%, in rough agreement with our 2D prolate ejecta model with $\rho_{\rm pole}/\rho_{\rm eq}=$\,5. Choice ISP4 implies a high intrinsic polarization but a very weak depolarization within lines, which is not as well matched by our models; as noted by L00, such a high degree of ISP also strains the allowable values from reddening considerations.  We thus arrive at the conclusion that the SN 1998S spectropolarimetry likely suffered little ISP contamination; in fact, assuming zero ISP produces results in as good agreement with our model predictions as any other.

 As the data in the $q-u$ plane do not lie perfectly along a straight line (Fig.~\ref{fig_obs_q_u}), the ejecta associated with SN1998S must exhibit departures from axial symmetry. The small degree of polarization angle rotation across the broad emission lines is not presently captured by our models and is a limitation of the imposed axisymmetry in \longpol. Modeling of such features requires 3D polarized radiative-transfer and is left to future work.


\section{Conclusion}
\label{sect_conc}

We have presented an end-to-end modeling of RSG stars exploding inside an asymmetric CSM. The models are based on 1D radiation-hydrodynamics calculations with \heracles\ of the ejecta interaction with the CSM, the post-treatment of multiepoch snapshots with the 1D radiative transfer code \cmfgen, and finally the 2D polarized radiative transfer with \longpol. The main limitation of our work is the incomplete physical consistency since asymmetry is only introduced at the last modeling stage with \longpol. Another adjustment is the enforcement of homologous expansion, as currently required by \longpol, although we show that this modification is acceptable (see Section~\ref{sect_setup} and Appendix~\ref{appendix_cmfgen}). We select the interaction model r1w6b (and offshoots r1w6[a,c]), presented and confronted to SN\,2023ixf by \citet{jacobson_galan_23ixf_23}.

Adopting a simple, depth-independent, latitudinal scaling of the density and assuming suitable scaling relations for the opacities and emissivities computed with \cmfgen\ for the 1D model r1w6b, we model the spectropolarimetry of 2D, prolate ejecta with a pole-to-equator density ratio of five. In those H-rich, ionized environments where the electron-scattering opacity dominates everywhere away from line cores, we find that the polarization is essentially constant with wavelength apart from funnel-shaped depolarizations across lines. The lower polarization across lines follows from the formation of lines exterior to the continuum, and thus at lower electron-scattering optical depth, but the additional depolarization is caused by the disappearance of continuum photons that wander in $\lambda$-space too close to highly-absorbing line cores. Our simulations with a fixed, depth-independent asymmetry, suggest the overall polarization should remain constant for several days in SNe II-P interacting with CSM before dropping, and flipping sign, as the shock emerges from the CSM. For a fixed CSM density, models with more compact/extended CSM exhibit a lower/greater maximum polarization with a more/less rapid evolution.

Our finding that the narrow line cores should exhibit some level of depolarization is used to set constraints on the ISP of SN\,1998S, leading to the conclusion that the intrinsic polarization of SN\,1998S is of $\sim$\,2\,\%. Our models replicate the wavelength independence of the continuum polarization, with the funnel-shaped depolarization across lines. This similarity lends support to the essential ansatz of our models and the numerical approach. As discussed in Section~\ref{sect_res} (see also results for models r1w6[a,c] in Appendix~\ref{sect_appendix_figs}), tweaks to a few input parameters (e.g., $R_{\rm CSM}$, $\rho_{\rm pole}/\rho_{\rm eq}$, etc.) could in principle be made in order to provide better fits to the actual levels of (even temporally changing) continuum polarization for objects where the ISP is more tightly constrained than is the case for SN\,1998S. Overall, our methodology extended here to include spectropolarimetry, offers a means to model and constrain the properties of the growing sample of SNe II-P/CSM  like 2023ixf and better understand the origin of pre-SN mass loss.

\begin{acknowledgements}
This work was supported by the ``Programme National Hautes Energies'' of CNRS/INSU co-funded by CEA and CNES. This work was granted access to the HPC resources of TGCC under the allocation 2023 -- A0150410554 on Irene-Rome made by GENCI, France. D.C.L. acknowledges support from NSF grant AST-2010001, under which part of this research was carried out. D.J.H. gratefully acknowledges support through NASA astro-physical theory grant 80NSSC20K0524.
\end{acknowledgements}



\appendix

\section{The assumption of homologous expansion in \cmfgen\ simulations of interacting supernovae}
\label{appendix_cmfgen}

\begin{figure}
\centering
\includegraphics[width=\hsize]{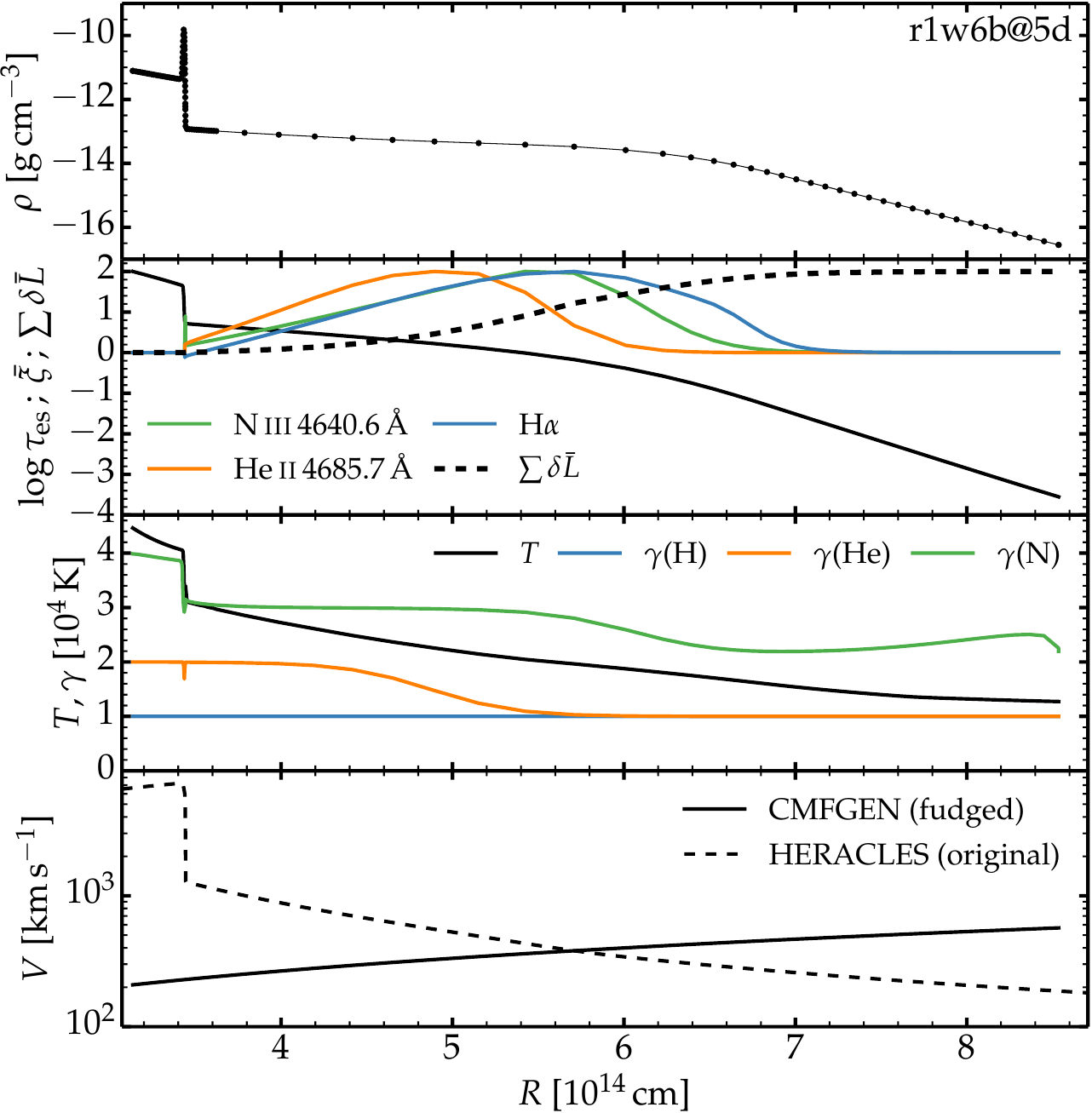}
\caption{Illustration of ejecta and radiative properties of model r1w6b at 5\,d after explosion, as computed by \heracles\ and \cmfgen. From top to bottom, we show the variation with radius of the mass density (dots indicate the location of the hundred grid points in the \cmfgen\ simulation), the electron-scattering optical together with the regions of formation of H$\alpha$, He\two\,4685.7\,\AA, and N\three\,4640.6\,\AA, the temperature together with the H, He, and N ionization (a value of one means a species once ionized), and the original, nonmonotonic velocity from the \heracles\ simulation (dashed line) together with the fudged, homologous velocity adopted for the \cmfgen\ calculation (solid line). [See discussion in Section~\ref{sect_setup}.]}
\label{fig_init}
\end{figure}

We show some results from the \cmfgen\ calculation assuming homologous expansion in Fig.~\ref{fig_init} and using the model r1w6b ($R_{\rm CSM}=$\,8$\times$\,10$^{14}$\,cm) -- this model was found by \citet{jacobson_galan_23ixf_23} to yield a satisfactory match to the photometric and spectroscopic evolution of SN\,2023ixf, a close analog of SN\,1998S (model r1w6b derives from model r1w6 of \citet{d17_13fs}, which has been extensively used in the SN community). The time is 5\,d after first detection.

Looking at the velocity structure first, we see that the original, nonmonotonic velocity and the new (``fudged''), homologous velocity (bottom panel of Fig.~\ref{fig_init}) differ significantly. By assuming homologous expansion, the fast inner regions are now the slowest while the originally slow outer CSM regions are now the fastest. The unadulterated velocities differ from the initial velocities because of the radiative acceleration of the unshocked CSM (see simulation results for this phenomenon in  \citealt{d17_13fs}; see also \citealt{chugai_98S_02}). Here, in the \heracles\ calculation, the entire CSM is predicted to move at a velocity greater than 200\,\kms, and as fast as 300\,\kms\ at the photosphere location at 5\,d. Such relatively large velocities suggest that a resolution of 300\,\kms\ is not so bad for spectropolarimetric observations of interacting SNe (as obtained by L00 for SN\,1998S).

These offsets have, however, a moderate impact for what concerns us. First, the inner, fast moving layers are at high optical depth and thus contribute negligibly to the emergent flux (both are shown in the second panel from top in Fig.~\ref{fig_init}). The outer CSM regions are of low density and low optical depth and will thus have a weak impact, essentially limited to the narrow emission line cores. Because these velocities are still small, they are typically unresolved in spectropolarimetric observations -- our models will simply overestimate the width of this narrow, line core emission by a factor of about two. The bulk of the spectrum forms between optical depth 0.1 and 10 (i.e., around the photosphere; see quantity $\sum \delta \bar{L}$ in Fig.~\ref{fig_init}, second panel from the top), where the original and the modified velocities are similar by design. Electron scattering being the dominant line broadening mechanism at such times and in those regions, and since we are not concerned here with information at the 100\,\kms\ scale only available in high-resolution spectra, this slight change in material velocity is unimportant.

This adjustment of the original nonmonotonic velocity into a homologous flow leads to different velocities at different epochs or in different models. While we set $V=R/t$ and $t=R_{\rm phot} / V_{\rm phot}$ in all cases, the quantities $R_{\rm phot}$ and $V_{\rm phot}$ are specific of each \heracles\ snapshot for each model studied. Obviously, this ``homologous'' time does not correspond to any physical time for either the ejecta, the CSM, or the interacting SN, but this time plays no role in the steady-state radiative transfer to be performed.

\section{Polarized radiative transfer with \longpol}
\label{sect_nomenclature}

For completeness, we summarize the nomenclature and sign conventions adopted in \longpol\ and also presented by \citet{DH11_pol}.  We assume that the polarization is produced by electron scattering. The scattering of electromagnetic radiation by electrons is described by the dipole or Rayleigh scattering phase matrix. To describe the ``observed'' model polarization we adopt the Stokes parameters  $I$, $Q$, $U$, and $V$ \citep{Cha60_rad_trans}. Since we are dealing with electron scattering, the polarization is linear and the $V$ Stokes parameter is identically zero. For clarity, $I_Q$ and $I_U$ refer to the polarization of the specific intensity, and $F_Q$ and $F_U$ refer to the polarization of the observed flux.

For consistency with the  earlier work of \cite{hillier_94, hillier_96} we choose a right-handed set of unit vectors $(\zeta_X,  \zeta_Y, \zeta_W)$. Without loss of generality the axisymmetric source is centered at the origin of the coordinate system with its symmetry axis lying along $\zeta_W$, $\zeta_Y$ is in the plane of the sky, and the observer is located in the XW plane.

We take $F_Q$ to be positive when the polarization is parallel to the symmetry axis (or more correctly parallel to the projection of the symmetry axis on the sky), and negative when it is perpendicular to it. With our choice of coordinate system, and since the SN ejecta are left-right symmetric about the XW plane,  $F_U$ is zero by construction. This must be the case since symmetry requires that the polarization can only be parallel, or perpendicular to, the axis of symmetry. For a spherical source, $F_Q$ is also identically zero.

$I(\ip,\delta)$, $I_Q(\ip,\delta)$ and $I_U(\ip,\delta)$ refer to the observed intensities on the plane of the sky. $I_Q$ is positive when the polarization is parallel to the radius vector, and negative when it is perpendicular. In the plane  of the sky we define a set of polar coordinates ($\ip,\delta$)  with the angle $\delta$  measured anti-clockwise from $\zeta_Y$. The polar coordinate, $\ip$, can also be thought of as the impact parameter of an observer's ray. We also use the axes defined by the polar coordinate system to describe the polarization. $F_I$ is obtained from $I(\ip,\delta)$ using

\begin{equation}
F_I= {2 \over d^2}  \int_0^{\ip_{\rm max}} \int_{-\pi/2}^{\pi/2}  \, I(\ip,\delta) dA \, ,
\end{equation}

\noindent
where $dA=\ip d\delta d\ip$. Since $\zeta_\ip$ is rotated by an angle $\delta$ anticlockwise from $\zeta_Y$,
$F_Q$ is given by

\begin{equation}
F_Q = {-2 \over d^2} \int_0^{\ip_{\rm max}}  \int_{-\pi/2}^{\pi/2}
\left[ I_Q(\ip,\delta)  \cos 2\delta +  I_U(\ip,\delta)   \sin 2\delta \right] \, dA \,.
\end{equation}

\noindent
In a spherical system, $I_Q$ is independent of $\delta$, and $I_U$ is identically zero.

\noindent
The percentage polarization $P_\lambda$ is defined as $100  | F_Q \, \big / F_I|$, where we have dropped
the $\lambda$ subscript of the fluxes for clarity.

In this paper and for brevity, we report mostly on the maximum polarization obtained for a 90-deg inclination angle $i$. The variation with $i$ has been discussed in previous studies and may deviate from a simple $\sin^2 i$ dependence if optical-depth effects are present \citep{dessart_pol_blob_21,dessart_12aw_21}.

\section{Additional models and illustrations}
\label{sect_appendix_figs}

\begin{figure}
\centering
\includegraphics[width=\hsize]{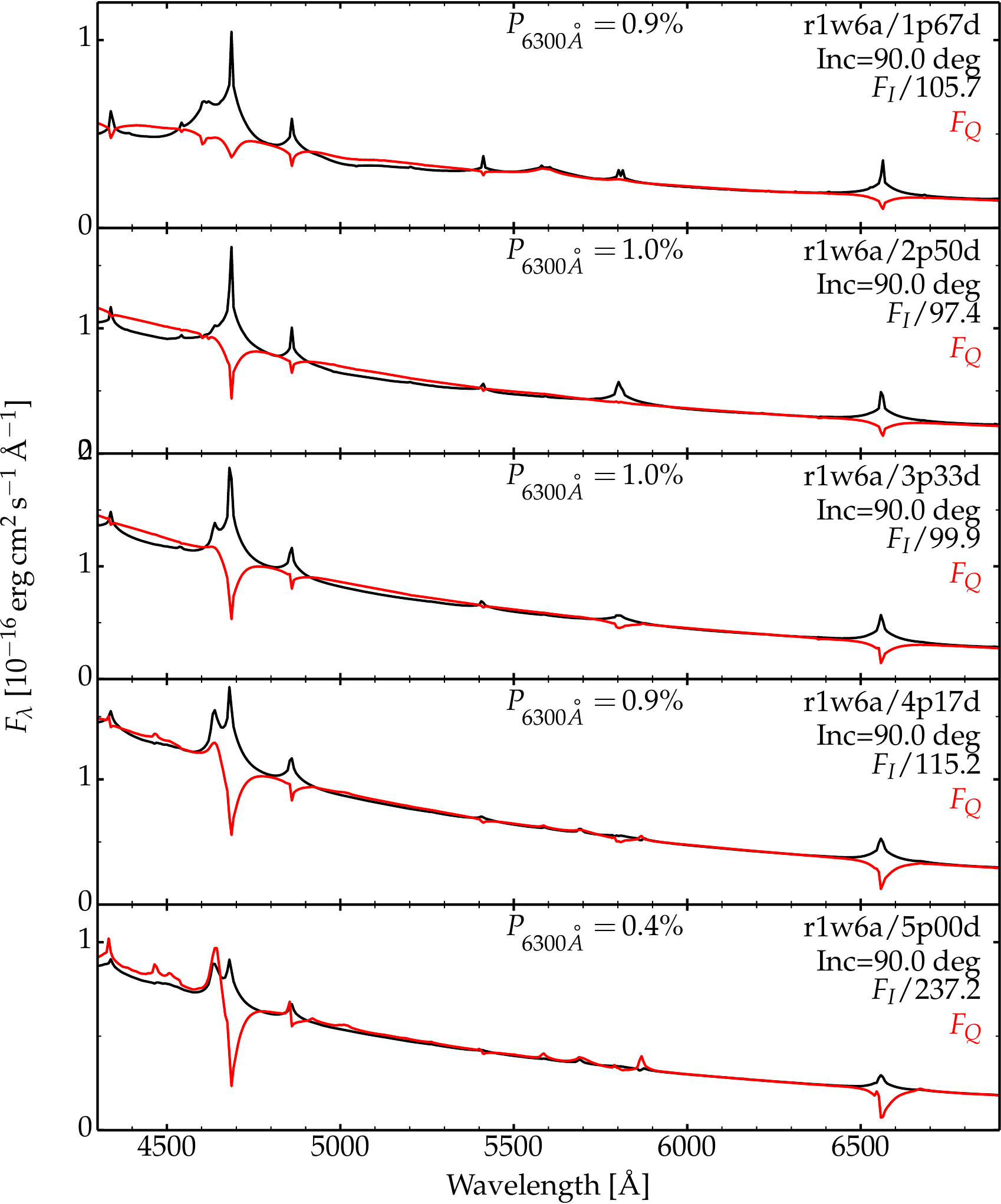}
\caption{Same as Fig.~\ref{fig_r1w6b_evol} but now for model r1w6a and from 1.67 to 5.00\,d.}
\label{fig_r1w6a_evol}
\end{figure}

\begin{figure}
\centering
\includegraphics[width=\hsize]{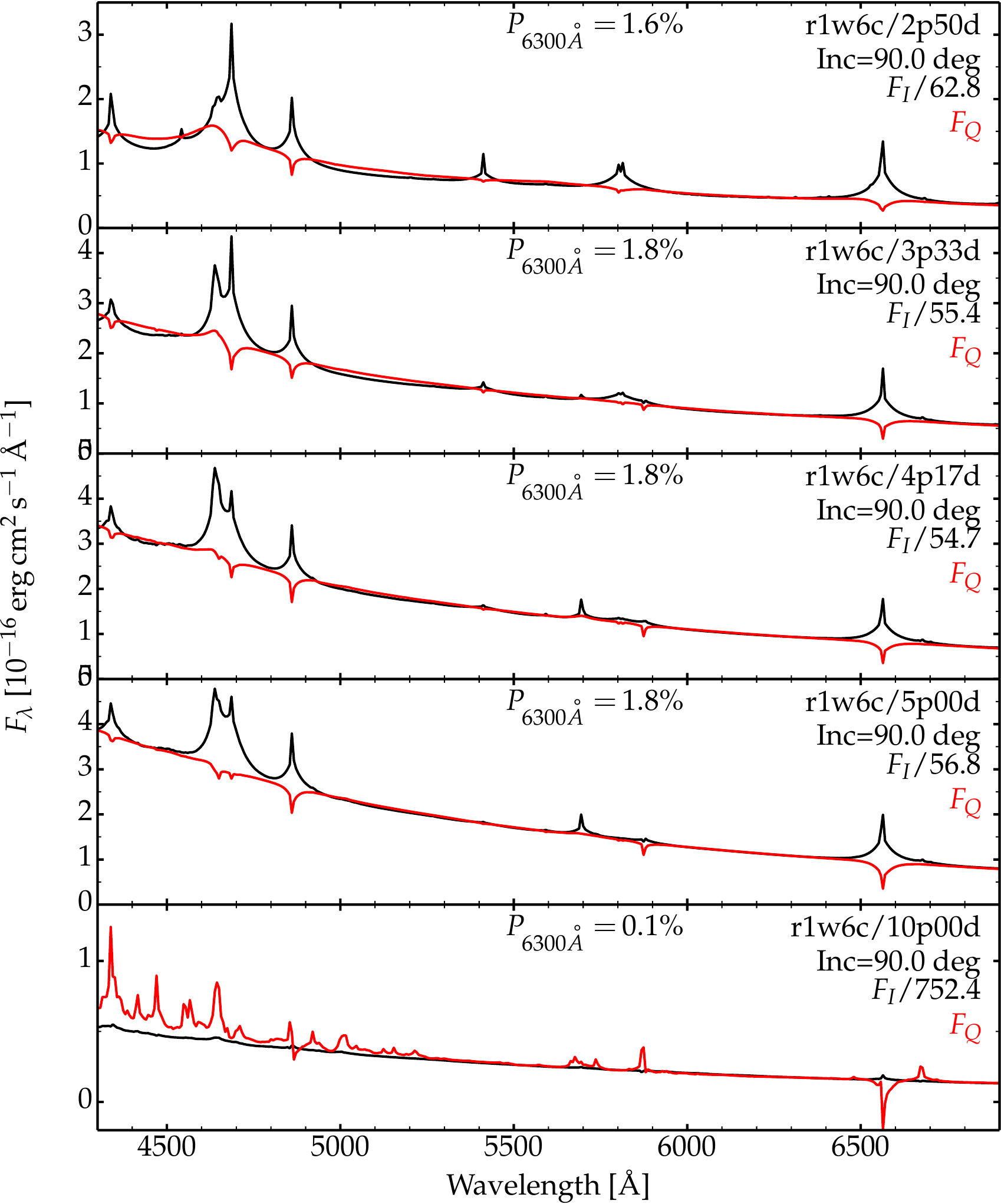}
\caption{Same as Fig.~\ref{fig_r1w6b_evol} but now for model r1w6c}
\label{fig_r1w6c_evol}
\end{figure}

\end{document}